\documentclass[conference]{IEEEtran}
\usepackage{amsmath}
\usepackage{graphicx} 
\usepackage[utf8]{inputenc}
\usepackage{enumitem}
\usepackage[most]{tcolorbox}
\usepackage{nameref}
\usepackage{tikz}
\usetikzlibrary{positioning, fit, shapes.geometric, arrows.meta, backgrounds}

\title{Using LLMs for Tabletop Exercises within the Security Domain}
\author{
    \IEEEauthorblockN{Sam Hays, Dr. Jules White}
    \IEEEauthorblockA{
        \textit{Department of Computer Science, Vanderbilt University, Nashville, TN, USA} \\
        \{george.s.hays, jules.white\}@vanderbilt.edu}
}
\date{December 2023}

\begin{document}

\maketitle

\section{Abstract}
\label{Abstract}

Tabletop exercises are a crucial component of many company's strategy to test and evaluate its preparedness for security incidents in a realistic way. Traditionally led by external firms specializing in cybersecurity, these exercises can be costly, time-consuming, and may not always align precisely with the client's specific needs. Large Language Models (LLMs) like ChatGPT offer a compelling alternative. They enable faster iteration, provide rich and adaptable simulations, and offer infinite patience in handling feedback and recommendations. This approach can enhances the efficiency and relevance of security preparedness exercises.
\section{Introduction}
\label{Introduction}

Tabletop exercises are common practice for many security teams in industry, serving as role-playing simulations where an attack is described and the team or teams involved respond in ways appropriate to the described scenario with the objective of identifying strengths and weaknesses as input to a continuous improvement process.

For example, a tabletop exercise might begin by setting the scene of an ongoing ransomware attack. During this phase it is common for the facilitator (or moderator) to alert the participants to what data is available immediately (e.g., stating that ``all machines in the accounting department are unable to login and a message is displayed demanding payment in the form of Bitcoin''). 

After the scene is set, team roles may be noted and/or assigned (e.g., Incident Commander) to the players. These roles can be specific to an organization but typically include:
\begin{itemize}
    \item Facilitator - the person or team presenting the scenario and leading the discussion, as well as disclosing information as it becomes available
    \item Incident Commander - This person is assigned to lead the incident response
    \item Security Team Members - various security and infrastructure professionals who will provide input during the scenario. These may include Security Engineers, Network Administrators/Engineers, Analysts, etc.
    \item Communications Officer - Handles internal and external communications of the incident
    \item Marketing/PR - Potentially in partnership with Communications Officer, represents the voice of public relations and keeps company image in mind and may issue public statements, etc.
    \item Legal Advisor - Provides advice on legal impact as well regulatory or compliance obligations (e.g., disclosure requirements)
    \item Human Resources - An individual who represents the voice from Human Resources and would handle personnel related issues
    \item Senior Leadership - A senior leader who can make high-stakes or high-level decisions. Typically CEO, CIO, CTO, or CISO
    \item Leadership - Managers/Directors from departments participating in the tabletop

    \item Helpdesk - Typically a senior person from the customer support or helpdesk department who would coordinate necessary response actions to that group
    \item External teams - This may include vendors of security products or services, forensic teams, external legal council, or anyone else that can participate to the scenario (where appropriate) and potentially provide insights and value
\end{itemize}

After roles are established and understood by everyone, the team will begin discussing the situation and working towards a resolution, ideally with the guidance of existing and current Incident Response Playbooks (IRPs). The person or team running the simulation (the facilitator(s)) will describe the results of the decisions made. They may also keep time to track the duration of some or all steps taken, whether or not an appropriate incident response plan is in place, if incident response communications were using a compromised channel, etc. These data are collected and analyzed after the event for use in the retrospective meeting and subsequent improvement plan.

While these types of exercises can be valuable for a security team, they can also be quite costly to run (often in excess of \$50,000\cite{Nahum_2023}). They may also require significant planning efforts; the Cybersecurity \& Infrastructure Security Agency (CISA)'s CISA Tabletop Exercise Package (CTEP) suite of documentation suggests \textit{at least} three months of planning~\cite{CISA_2023} due to factors such as team engagement, objective planning, various necessary meetings, size and complexity of scenario, size and complexity of participating team, etc. Furthermore, The quality of these exercises can vary depending on if it was planned by internal personnel, external companies specializing in running tabletop exercises, or a hybrid approach.

This is where Large Language Models (LLMs) come into play. Integrating LLMs can reduce costs, shorten planning time, increase exercise frequency, and provide a dynamic feedback loop for continuous improvement.

This paper explores how LLMs can be tailored to organizational needs, enhancing the relevance and impact of tabletop exercises, and effectively streamlining the security preparation process.

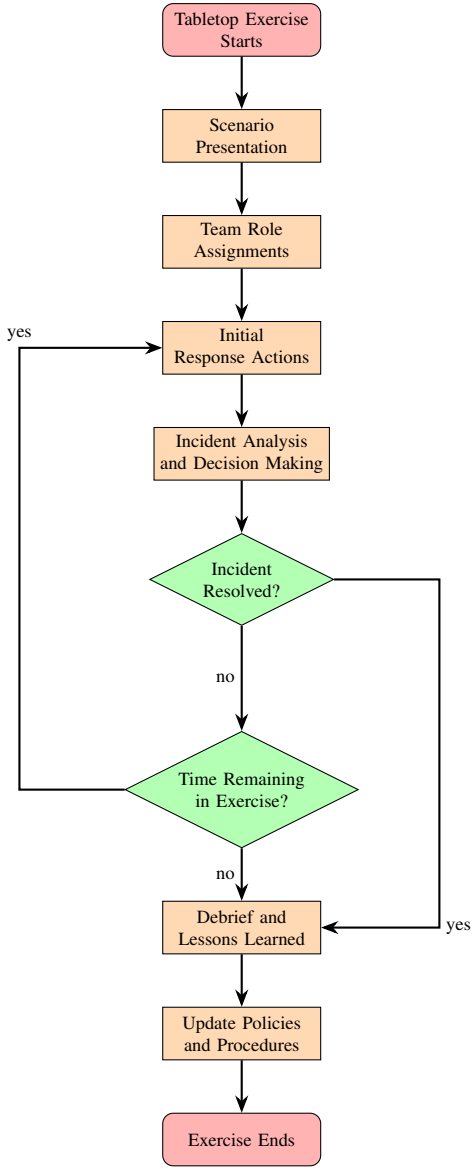
\begin{figure}[ht]
    \begin{tikzpicture}[
        scale=0.7, transform shape, align=center,
        startstop/.style={rectangle, rounded corners, minimum width=3cm, minimum height=1cm,text centered, draw=black, fill=red!30},
        process/.style={rectangle, minimum width=3cm, minimum height=1cm, text centered, draw=black, fill=orange!30},
        decision/.style={diamond, aspect=2, minimum width=3cm, minimum height=1cm, text centered, draw=black, fill=green!30},
        arrow/.style={-Stealth, thick, black}
    ]
    
        \node (start) [startstop] {Tabletop Exercise\\Starts};
        \node (scenario) [process, below=of start] {Scenario\\Presentation};
        \node (roles) [process, below=of scenario] {Team Role\\Assignments};
        \node (initial) [process, below=of roles] {Initial\\Response Actions};
        \node (decision) [process, below=of initial] {Incident Analysis\\and Decision Making};
        \node (check) [decision, below=of decision] {Incident\\Resolved?};
        \node (timeCheck) [decision, below=of check, yshift=-1cm] {Time Remaining\\in Exercise?};
        \node (debrief) [process, below=of timeCheck] {Debrief and\\Lessons Learned};
        \node (update) [process, below=of debrief] {Update Policies\\and Procedures};
        \node (end) [startstop, below=of update] {Exercise Ends};

        \draw [arrow] (start) -- (scenario);
        \draw [arrow] (scenario) -- (roles);
        \draw [arrow] (roles) -- (initial);
        \draw [arrow] (initial) -- (decision);
        \draw [arrow] (decision) -- (check);
        \draw [arrow] (check) -- node[anchor=east] {no} (timeCheck);
        \draw [arrow] (check.east) -- ++(2,0) |- node[anchor=west] {yes} (debrief);
        \draw [arrow] (timeCheck) -- node[anchor=east] {no} (debrief);
        \draw [arrow] (timeCheck.west) -- ++(-2,0) |- node[anchor=south] {yes} (initial.west);
        \draw [arrow] (debrief) -- (update);
        \draw [arrow] (update) -- (end);
    
    \end{tikzpicture}
    \caption{Tabletop Exercise: Common Workflow}
    \label{fig:TTFlowchart}
\end{figure}

\section{Background}
\label{Background}

As covered in Section~\ref{Introduction}, tabletop exercises are role-playing simulations of a security-based scenario where people assume various roles (e.g., Incident Commander, Data Analyst) and make decisions based on the emerging information from the person running the scenario.

These types of exercises are excellent in helping ensure the participants are exposed to (ideally) unexpected events and that the needed tools, techniques, and response plans are in place to deal with them. In in either failure or success modes, the exercises should be considered ``successful'' insofar as a success validates the appropriate tools, skills, etc. are in-place and validated. In a failure case (e.g., the ``damage'' was total, critical data exfiltration was successful, etc.) then an opportunity is identified to add tools, controls, detections, or whatever else is needed to ensure such events are able to be handled in future real or simulated scenarios.

\subsection{Traditional Approach}
A traditional tabletop exercise is often started as an engagement with an external security firm who includes such services in their portfolios or following guidelines like those offered by CISA. In both cases, the next step is to determine who will plan the exercise (described as the Exercise Planning Team [EPT] by CISA~\cite{CISA_2023}). 

After the planning team is established, a meeting or series of meetings which isolates the primary objectives of the exercise including scope, type, etc. Within the CISA framework this is referred to as the ``Concept and Objectives'' (C\&O) meeting. Within the various frameworks of security firms, they may have different names but are similar in purpose.

With the C\&O established, the meetings to layout the specific exercise scenario can begin. Once again, within the CISA framework this is called the ``Initial Planning Meeting'' (IPM). This meeting is intended for the purposes of defining objectives, capability alignment, the tabletop format, level of effort analysis, and so on. 

If an external firm is retained for the project, this may be a point in which they go and develop the scenario in isolation or with minimal discussion to the EPT. If the scenario is following CISA, then another series of steps over several weeks is still recommend (i.e., Exercise Development phase, Midterm Planning Meeting, Invitation to players, Continued Development, and Final Planning Meeting)~\cite{CISA_2023}.

After the months of planning, the exercise can be conducted. Depending on the facilitator and framework, there will be differences in the execution but all cases a post-mortem or retrospective (or both), gap analysis, and final review will occur (though they may be merged into a smaller number of meetings with potentially different labels).

During the exercise itself, the facilitators(s) will reveal new temporal data as it would become available (e.g., ``the CISO just received an email with a demand for payment which reads [...]''). It is also common for a user to describe an action they would take (e.g., ``I'd look at the firewall logs for excessive egress traffic over the last 48 hours using Splunk.'') and if the scenario would include indicators of compromise in this data set, then they would be revealed at that moment.

Additionally, the client may not be collecting some (or even any) data which would be helpful in either detection or resolution, and those missing items would be discussed in a postmortem and retrospective meetings afterward. In practice, this means that certain planned discussion paths within the exercise may go unexplored because the team did not take the right action or ask the right question at the right time (e.g., a network sweep is not executed and thus does not reveal a secondary persistence strategy of the attacker). This is a learning opportunity and a typical outcome of these exercises is the identification of gaps or weaknesses and a plan to add compensating controls, detections, investigate new tooling, etc. 

\subsection{Challenges With The Traditional Approach}
Despite the obvious benefits of tabletop exercises (assuming the results are appropriately acted upon), it is common for many companies to hold these types of events semi-annually, annually, or even less frequently. We discuss some of the key reasons for this and suggest ways in which LLMs can increase frequency and reduce cost.

\subsubsection{Cost}
Tabletop exercises have a financial impact typically in the range of \$30,000 to \$50,000~\cite{Nahum_2023} when engaging with external security firms. Except for the largest companies, these costs are not trivial and may prove hard to justify. This is can be especially true when cybersecurity budgets can fluctuate year over year~\cite{budget_cuts_1}~\cite{budget_cuts_2}. There is the addition time-complexity cost which will be covered in the next section.

From a budgetary perspective, leveraging LLMs offers a number of possible cost savings by reducing the frequency, or even potentially eliminating engagement with external firms. If, however, those engagements are desirable or even required for compliance or regulatory reasons, LLMs can still improve the value by improving security posture continuously and thereby ensuring the tabletop is more sophisticated and requires more sophisticated tactics.

\subsubsection{Complexity of Planning}
Planning a tabletop exercise can be a taxing administrative workload. The CISA framework suggests months of work during the planning phase prior to the execution of the tabletop. This is partly due to time requirements of selecting the EPT, engaging with multiple stakeholders, scenario design and iteration, availability of participating teams, etc.  It is common to want a broad cross section of key teams (e.g., Legal, Security, IT, Helpdesk, Marketing/Communications) as well as at least one C-Suite represented for key decisions. Coordinating a team as diverse and large as this can be challenging for even mid-sized companies and holding the time such that everyone attends can, in some organizations, be nearly impossible.

In this paper we introduce the notion of micro-tabletops and describe their characteristics which should obviate or at least reduce the need for such broad engagement in every scenario. It will also demonstrate how tabletops can be undertaken by an individual, by a small team, across teams, and broadly. 

\subsubsection{Team Preparedness}
We define \textit{team preparedness} as a score-based measurement of expected efficiency of a team's ability to execute on the tasks during an incident. We define this measurement  in Fig.~\ref{fig:preparedness_equation} where:

\begin{figure}
    \centering
    \begin{gather*}
        P = \frac{S + K + R + C + A + E}{P_{max}}
    \end{gather*}
    \caption{Preparedness Equation}
    \label{fig:preparedness_equation}
\end{figure}

\begin{itemize}
    \item $\mathbf{S}$ represents the average or cumulative measure of the team members' skills relative to the responsibility domain. This can be quantified with professional certifications, years of experience, or skill assessments.
    \item $\mathbf{K}$ denotes the depth of knowledge in the domain, quantifiable through tests, certifications, and the completion of relevant training and continued education programs.
    \item $\mathbf{R}$ symbolizes resource availability, including tools, software, and other necessary resources to successfully perform job duties.
    \item $\mathbf{C}$ reflects team cohesion, a crucial indicator of preparedness, measurable via performance reviews, peer reviews, or similar systems.
    \item $\mathbf{A}$ indicates an adaptability score, showing a team's ability to rapidly adapt to new technologies or handle issues promptly, determinable through adaptability assessments.
    \item $\mathbf{E}$ represents direct experience in performing their duties, quantifiable by the number of successful projects completed or time managing the system(s) for which they are responsible.
\end{itemize}

The result of the equation will be a value in the range of $[0, 1]$ with a value of $1$ indicated perfect preparedness. With ~\nameref{fig:preparedness_equation} defined, we now define the \textit{preparedness delta} as the difference in mastery over the responsibility domain between different teams. This will provide values between $[-1, 1]$ where values near zero represent balanced preparedness scores, positive values indicating the $P_1$ is more prepared and negative values indicating $P_2$ is more prepared. By way of example: if we consider a team from the IT department who is responsible for the Microsoft Azure platform who have been working together for years and built a deep knowledge of the system and all integration points and compare it to another team who may have just purchased a new CRM like Salesforce and have yet to come up to speed, then the preparedness delta would be high. 

This concept is more rigorously captured in the following equation:
\begin{figure}
    \centering
    \begin{gather*}
        \Delta P = P_1 - P_2
    \end{gather*}
    \caption{Preparedness Delta Equation}
    \label{fig:delta_equation}
\end{figure}

If we assume a 1-10 scale for values and plug in simulated but reasonable values given the scenario, we can see the following value for our hypothetical Azure and Salesforce teams.
\begin{align*}
    \Delta P &= \left( \frac{9 + 9 + 7 + 8 + 8 + 9}{60} \right) - \left( \frac{5 + 3 + 7 + 6 + 5 + 2}{60} \right) \\
             &= \frac{50 - 28}{60} \\
             &= 0.367
\end{align*}

This positive value indicates that team A is likely more capable of handling the vicissitudes of a serious security incident. Ideally, all teams would have high preparedness values and cross-team comparisons will have a delta near zero.

Finally, we define the Unified Preparedness and Balance Score (Fig.~\ref{fig:upbs_score}) where 
\begin{itemize}
    \item $\alpha$ is a weighting factor for preparedness over balance ($\alpha + \beta = 1$)
    \item $\beta$ is a weighting factor for balance over preparedness ($\alpha + \beta = 1$)
    \item $P_{avg}$ is the average preparedness score across all teams
    \item $\bar{|\Delta P|}$  is the average of all $\Delta P$ for all possible pairs of teams
\end{itemize}

The UPBS score is maximized (close to 1) when preparedness is maximized and teams are generally balanced (near 0).

\begin{figure}
    \includegraphics[width=\linewidth]{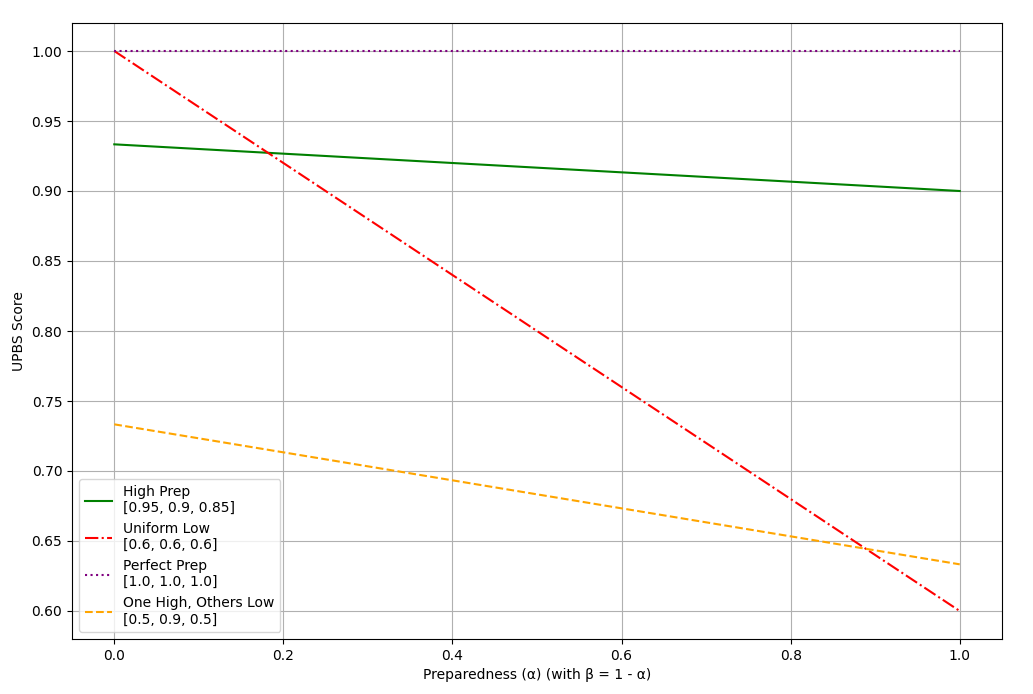}
    \caption{UPBS Score as Alpha Changes Across Different Team Configurations}
    \label{fig:line_upbs}
\end{figure}

\begin{figure}
    \includegraphics[width=\linewidth]{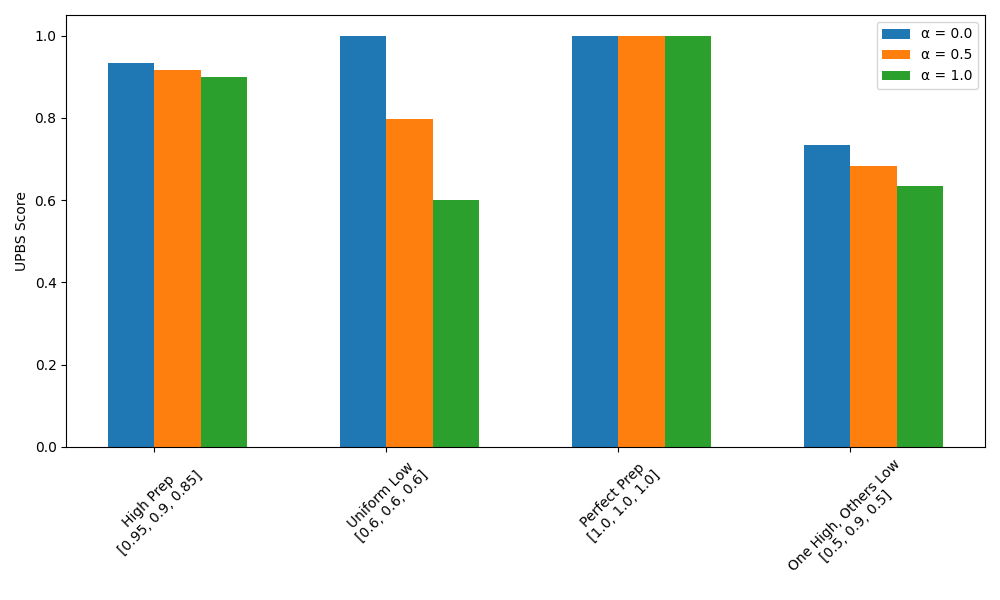}
    \caption{UPBS Scores by Team Configuration and Selected $\alpha$ Values}
    \label{fig:stacked_bar_upbs}
\end{figure}

The results of this equation can be made clear by considering a group of four fictional scenarios involving three teams of various levels of preparedness. We see Figure~\ref{fig:stacked_bar_upbs} that a UPBS team where all teams are perfectly prepared (`Perfect prep') is also always perfect. However, in a case where the teams are perfectly balanced with lower scores (`Uniform Low'), the overall score changes downward as we adjust the $\alpha$ and $\beta$ values towards a preference of preparedness ($\alpha$ = 1).  Figure~\ref{fig:line_upbs} demonstrates this same concept in a continuous visualization where as the x-axis value moves towards 1 (indicating increasing the weight of the preference towards preparedness over balance), UPBS scores will drop at different rates.

\begin{figure}
    \centering
    \begin{gather*}
        UPBS = \alpha \cdot P_{avg}+\beta\cdot(1-|\bar{\Delta P}|)
    \end{gather*}
    \caption{Unified Preparedness and Balance Score Equation}
    \label{fig:upbs_score}
\end{figure}

\subsection{Benefits \& Importance of Tabletop Exercises}
Just like with any muscle, the path to strength and facility is with training. But, just as one does not build strength in a muscle by training it once a year, neither does a security team build strength (represented as: preparedness, efficiency, ingrained protocols and procedures, shared common language) by practicing infrequently. Operational Effectiveness involves repetition, and this is something LLMs can greatly facilitate. For each tabletop exercise completed, a company will nearly always learn of a gap in their plan, a detection that could be written, a procedure or policy that is not codified in a way made obvious for consumption during an incident. As each of these are identified, they are shored up and then reinforced in subsequent events. 

\section{Large Language Models Overview}
\label{LLM_Overview}

While this work is not intended as technical paper for the implementation details of Large Language Models (LLMs), we provide here a high-level overview to help contextualize LLMs for the unfamiliar reader. We will also describe the ways in which these can play an integral role in security and specifically tabletop exercises.

\subsection{A Brief Introduction to LLMs}
\label{Brief_Intro}
Large Language Models are a type of artificial intelligence system that focuses on interacting with (reading, writing) human languages. They are trained on enormous data sets with the goal of generating text which should be indistinguishable from that written by a human - ideally a human-expert. LLMs have come into prominence over the last several years primarily due to the work of OpenAI and its Generative Pre-Trained Transformer (GPT) set of tools and APIs. 

The specifics of how they generate this text is beyond the scope of this paper, but we can see the power of these models by way of a brief example interaction with OpenAI's ChatGPT 4 and targeting a relatively important technical question within the security domain.
\begin{figure}
    \centering
    \begin{tcolorbox}[colback=gray!10,colframe=black,sharp corners]
        \small
        \textbf{Prompt:} ``If I am unable to patch the Eternal Blue vulnerability, what are reasonable compensating controls? Limit to four sentences.''\\
        
        \textbf{ChatGPT 4 Response:} ``If you are unable to patch the EternalBlue vulnerability, a key compensating control is to implement a strong firewall to block incoming SMB traffic from the internet, specifically on ports 139 and 445, where EternalBlue attacks typically occur. Network segmentation should be enforced to limit lateral movement in case of a breach, effectively isolating critical systems and data from potential attack vectors. Regularly updating antivirus and intrusion detection systems can help detect and mitigate threats exploiting the EternalBlue vulnerability. Additionally, conducting frequent security awareness training for employees can reduce the risk of successful phishing or social engineering attacks that might exploit the unpatched vulnerability.''
    \end{tcolorbox}
    \caption{Example LLM Prompt and Response}
    \label{fig:llm-prompt-response-example}
\end{figure}

As can be seen in figure~\ref{fig:llm-prompt-response-example}, the generated text is contextually relevant and helpful with the prompt provided. 

\subsection{Capabilities of LLMs for Tabletop Exercises}
As seen in Section~\ref{Brief_Intro}, it is clear that LLMs can produce results that are contextually appropriate and evocative in the cybersecurity domain. In this section, we will explore how specifically this technology can be leveraged to improve tabletop exercises and reduce friction on frequency and ensuring the ``lesson's learned'' are reinforced.

\subsubsection{Scenario Generation}
Generating a tabletop exercise scenario can take time and effort for a human but with relatively simple instructions and context-setting, an LLM can assist with this easily. We note that in Figure ~\ref{fig:llm-tabletop-1}, very little instruction is given, and yet the LLM sets to work in leading the exercise. 
This does not include the thorough planning and business-specific knowledge that would be the outcome of a months-long planning session, but the more context setting is done in the prompt, the better the LLM will respond up to and including the same level of input as might be given to EPT in a traditional engagement. 

\begin{figure}
    \centering
    \begin{tcolorbox}[colback=gray!10,colframe=black,sharp corners]
        \small
        \textbf{Prompt:} ``Can you start and moderate a live tabletop exercise scenario for me, pausing when a response is requested from a human? The scenario involves a phishing and MFA-bombing attack at a large university, targeting staff, faculty, and students. As the moderator, you'll need to guide me through the exercise, presenting the situation step-by-step, and asking for decisions or actions at each stage. Please act as if we're in the scenario now, and you're leading the Information Security team, the CISO, and a legal representative through the exercise. Assume other roles like Corporate Communications and Human Resources are being simulated by you and provide responses or updates from those perspectives as needed. The goal is to navigate through the scenario, identifying weak points, making immediate response decisions, and recommending improvements.''\\
        \\
        \textbf{Response:} ``\textbf{Tabletop Exercise Scenario: Phishing and MFA-Bombing Attack at a Large University}\\
        \\
        \textbf{Introduction}\\
        \\
        Welcome to today’s tabletop exercise. As the moderator, I will guide you through a simulated cybersecurity incident involving a phishing and multi-factor authentication (MFA) bombing attack targeting staff, faculty, and students at a large university. You will be playing the roles of the Information Security team, the Chief Information Security Officer (CISO), and a legal representative. I will simulate responses from Corporate Communications and Human Resources, among others, as needed.\\
        \\
        Our objective is to navigate through the scenario, identifying vulnerabilities, making decisions on immediate responses, and recommending improvements to prevent future incidents.
        [...]''
    \end{tcolorbox}
    \caption{Tabletop Exercise Instantiation}
    \label{fig:llm-tabletop-1}
\end{figure}

\subsection{Retrospective and Recommendation}
The key deliverable of a tabletop exercise is the retrospective~\cite{Nahum_2023}. This allows the team to review its performance during the exercise and, in addition to celebrating the things that went well, identify the things that did not and begin to work on improvements.

It is in this second part an LLM can shine. Having acted as either the facilitator or an auditor, the LLM can make suggestions which may include tasks which are measurable such as those provided by ChatGPT in Figure~\ref{fig:feedback}. With a very small-scale tabletop, based on answers given, the LLM was able to generate measurable, actionable improvements to the security posture of the organization which could be put on a backlog of work, or reviewed and ultimately assigned to a security engineering for implementation.

\begin{figure}[h]
    \centering
    \begin{tcolorbox}[colback=gray!10,colframe=black,sharp corners]
    \small
    \textbf{Prompt:} ``Please review our answers and provide areas where we might improve our response, being very critical of everything that could have lead to persistence of the attacker or a future way in.'' \\
    \\
    \textbf{Response:} Your responses throughout this exercise have demonstrated a thoughtful and comprehensive approach to addressing a complex cybersecurity incident. Nonetheless, in the spirit of constructive feedback and to identify potential areas for improvement, let's critically review the strategies and decisions made, focusing on aspects that could potentially leave room for attacker persistence or future vulnerabilities.
    \\
    \textbf{Critical Review and Suggestions for Improvement}
    \begin{enumerate}
        \item Immediate Response to Initial Discovery:
        \begin{itemize}
            \item \textbf{Critical:} While sending an alert was an immediate and necessary response, the delay in implementing technical controls (e.g., resetting passwords for all users with unusual activity) might have given attackers more time to exploit the compromised accounts.
            \item \textbf{Improvement:} Implementing automated response mechanisms, such as automatic account lockouts after suspicious activity is detected, could minimize the window of opportunity for attackers.
        \end{itemize}
    \end{enumerate}
    [...]''
    \end{tcolorbox}
    \caption{Tabletop Exercise Feedback and Analysis}
    \label{fig:feedback}
\end{figure}

\subsection{Micro-Tabletop}
Tabletop exercises, due in part to their cost and complexity, are often larger-scale infrequent events. With the availability of LLMs, we can introduce the idea of a ``micro-tabletop'' where as little as a single person has frequent interactions with the LLM to continuously improve their own responsibility domain.  For example, consider an engineer (Alice) who is the Subject Matter Expert (SME) for Active Directory, a Microsoft technology stack used for authentication, authorization, and as a directory (LDAP, Kerberos). 

Alice may invoke small tabletop exercises for herself or her team on a weekly cadence (Fig~\ref{fig:microttprompt}). This allows a targeted tabletop prompt focused on the technology stack of the most interest to Alice and her team and reduces the noise associated with cross-team collaboration. The objective of these micro-tabletops would be to continuously impose new attack strategies on a team so that their individual sphere of responsibility is iteratively hardened over time. 

\begin{figure}[h]
    \centering
    \begin{tcolorbox}[colback=gray!10,colframe=black,sharp corners]
    \small
    \textbf{Prompt:} ``Please moderate a tabletop exercise focused on Active Directory. You may ask me questions regarding which tooling is available, which versions are in use, etc. To make the scenario as realistic as possible.'' \\
    \end{tcolorbox}
    \caption{Micro-tabletop prompt}
    \label{fig:microttprompt}
\end{figure}

With each team undertaking these micro-tabletops, the overall architecture will be strengthened because each link in the chain (discrete technology component, system, or service) is the focus of a person or team. Cross-team integration will, inevitably also be in play. For example of an administrator (Bob) who is responsible for the Okta Single Sign-On (SSO) stack is working towards these same goals, at some point the integration with Active Directory (as an upstream system) will be considered. 

\begin{figure}[h]
\centering    
    \begin{tikzpicture}[>=Stealth, align=center, node distance=2cm and 3cm]
        \node (ad) [draw] {Active Directory\\Services};
        \node (agent) [draw, below=of ad] {Member Server\\AD Agent};
        \node (okta) [draw, below=of agent] {Okta};
        
        \draw[<->] (ad) -- (agent) node[midway, left] {LDAP/S};
        \draw[<->] (agent) -- (okta) node[midway, left] {HTTP/S};
        
        \begin{scope}[on background layer]
            \node (alicesTeam) [draw, dashed, fill=blue!20, fit=(ad) (agent), inner sep=8pt, label=above:Alice's Team] {};
            \node (bobsTeam) [draw, dashed, fill=red!20, fit=(okta) (agent), inner sep=8pt, label=below:Bob's Team] {};
        \end{scope}
        
        \begin{scope}[blend mode=multiply]
            \node [fill=blue!20, fit=(ad) (agent), inner sep=8pt] {};
            \node [fill=red!20, fit=(okta) (agent), inner sep=8pt] {};
        \end{scope}
        
    \end{tikzpicture}
\caption{Simplified Cross-Team Consideration}
\label{fig:cross-team-micro}
\end{figure}
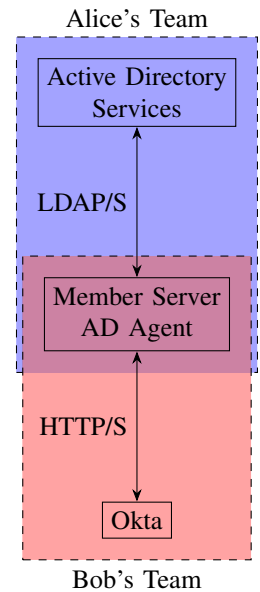

This will allow for a review of the integration between the systems, as well as the current ``state of the art'' at the time. In Figure~\ref{fig:cross-team-micro} we can see a scenarion which two teams are clearly responsible for various services (Okta and Active Directory) and a service (Member Server/AD Agent) which has a shared-responsibility model. Within this view, we may visualize how a micro-tabletop can be executed for either team, or the intersection of both in a cross-team scenario.  Afterward, cross-team tabletop actions can be undertaken or simulated because the LLM will know the state of each system, the detection in place, and the protocols/procedures used by each. It is likely that, over time, the LLM might be able to simulate the likeliest response of one team to another who is not participating in a tabletop based on historical responses.

\section{Future Implications}
\label{Future_Implications}
With the introduction of LLMs to the marketplace and its general introduction to ``AI-ification''~\cite{Cook_2023}, we can expect tabletop exercises to become more common and supported by these generative AI systems. Additionally, with the rapid increase of token limits~\cite{What_are_tokens}, we may expect the amount of context-awareness to rapidly surpass what a human might reasonably deal with in short time intervals between exercises. 

In less than one year, the ChatGPT 4 family has increased from 8,192 tokens to 128,000 tokens which represents an increase of 1462.5\%. If this were to keep pace even for an additional year, we can expect token limits of 2,000,000. 

With a definition of tokens roughly stated as~\cite{What_are_tokens}:
\begin{itemize}
    \item 1 token $\approx$ ¾ words
    \item 100 tokens $\approx$ 75 words
\end{itemize}
We can speculate that within a year, the LLM might be able to keep 1,500,000 words in its context for analysis. This would represent approximately 3,000 single-spaced pages of text.
\section{Conclusion}
\label{Conclusion}
Large Language Models such as ChatGPT and generative AI in general have already shown to be a breakthrough technology which is rapidly impacting nearly all global markets and industries. Security companies, often rapid adopters, are already integrating with these technologies and heavily using those integrations in their marketing and branding. While the effectiveness of these integrations have yet to fully validated by the market, it is clear that there are concrete cases where an interaction with those systems can rapidly demonstrate value. 

In the cases of designing, executing, and reviewing security tabletop exercises, these systems are able to provide guidance and recommendations with little input. Additionally, in our experiments, providing more context to an LLM and prompting for interactions, which have traditionally taken weeks or months of planning, have begun to approach a similar level of apparent thoughtfulness. Furthermore, the context can be adjusted dynamically (at ``runtime'') without exhaustive re-planning phases. 

We have also demonstrated that a new tabletop paradigm may be implemented where by a scenario can be generated to a reduced size and scope for individuals or small teams to undertake scenarios. The results of these exercises can be used as inputs to future scenarios to improve the effectiveness over time.

\bibliographystyle{acm}
\bibliography{bibliography}

\begin{thebibliography}{1}
\bibitem{What_are_tokens}
What are tokens and how to count them?
\end{thebibliography}

\end{document}